\documentclass[twocolumn,showpacs,preprintnumbers,amsmath,amssymb]{revtex4}
\usepackage{graphicx}
\usepackage{dcolumn}
\usepackage{bm}
\begin{document}

\title{THE SIMPLEST MODEL OF SPATIALLY DISTRIBUTED POPULATION WITH REASONABLE MIGRATION OF ORGANISMS}

\author{Michael G.Sadovsky}%
\email{msad@icm.krasn.ru}
\affiliation{%
Institute of biophysics of SD of RAS;\\ 660036 Russia, Krasnoyarsk, Akademgorodok.
}%

\begin{abstract}
The simplest model of a smart spatial redistribution of individuals is proposed. A
single-species population is considered, to be composed of two discrete subpopulations
inhabiting two stations; migration is a transfer between them. The migration is not
random and yields the maximization of net reproduction, with respect to the transaction
costs. The organisms are supposed to be globally informed. Discrete time model is
studied, since it shows all the features of a smart migrations, while the continuous time
case brings no new knowledge but the technical problems. Some properties of the model are
studied and discussed.
\end{abstract}

\pacs{87.23.Cc, 87.23.Kg, 87.23.-n}

\maketitle

\section{\label{introd}Introduction}
Modelling of the dynamics of biological takes start from the works by V.Volterra
\cite{volt,volt1} (see also \cite{lotka}). The approach follows, in brief, a chemical
kinetics, where reproduction, death and other types of interaction of organisms are
described with law of mass action. A spatially distributed population is modelled with
the ``reaction~--~diffusion'' system, where diffusion is supposed to describe the
redistribution of organisms over a space. Good coincidence of the solutions of such
``reaction~--~diffusion'' systems to observed dynamics of populations and communities
just masks the serious problem.

The point is that the diversity and abundance of possible (structurally stable) regimes
of such dynamic systems exceed drastically any really available family of trajectories of
real systems. Thus, one always is able to match a differential equation (or a system of
differential equation, either ordinary, or partial differential equations) to the dynamic
behavior of any really observed system. Doubtlessly, living organisms, even
microorganisms, differ drastically in their ``microscopic'' behavior from the chemical
substances and relevant chemical reactions.

Vito Volterra, the founder of the mathematical biology of populations understood pretty
well this discrepancy. Later, the comprehension of limitations of this chemical
methodology fell off. The situation is going worse when one tends to model a dynamics of
spatially distributed populations and communities. Basically, ``reaction -- diffusion''
systems make the basis for modelling of such spatially distributed entities
\cite{lotka,1,ber,ber3,xehi,alstad,murray,thieme,edel2,bern,turch,gi-co04}. Famous soviet
mathematician Andrew Kolmogorov studied in detail such type of equations \cite{kolmog}.

A methodology of modelling of spatially distributed populations based of the reaction --
diffusion system has the great discrepancy. It constrains significantly the individual
(behavioral) properties of organisms under consideration: one must suppose that the
organisms move over a space randomly and aimlessly \cite{ecol2003}. Obviously, such
assumption does not hold true (see, e.g. \cite{sci2000,insect,annals}), even for
microorganisms \cite{bolgar,zelkniga,levitt,ameba}. The assumption towards the idle
transfers of organisms in space is obviously less favorable for the higher organised
species.

Modelling methodology based on evolution principles is the way to pass over the
discrepancy mentioned above. This approach takes the origin from the evolution studies of
J.B.S.Haldane \cite{haldane}. This is the most general principle prescribing the way
biological systems evolve. In brief, it force to evolve a system toward the maximization
of net reproduction. This latter is an average number of {\sl per capita} descendants
survived at the course of a series of reproductions over an arbitrary long generation
line \cite{otbor1,otbor2,otbor3,otbor4}. Later, they found this principle to be even more
general, than just a biological one. Indeed, the principle holds true for any system
where the inheritance takes place \cite{rozonoer,zah-usp,zah-spek,lvov,ezer,na2}.

A consistent and rational implementation of this principle faces the problem of a lack of
knowledge of how the specific biological issues impact the survival of a species. In
turn, the question arises, what is an entity to be evolving? An ordinary answer on this
question is that the species is an evolving entity. Actually, the situation looks more
complicated; not discussing this problem in detail, further we shall follow this idea.
The principle formulated above yields the following rule for the model implemented below:
evolution optimality in spatial distribution of organisms is equivalent to the
maximization of (an average) net reproduction over space, with respect to the evolution
trade off for such redistribution. Some further details on this issue could be found in
\cite{otbor1,otbor4,na2,na4}.

\section{Basic model of the smart migration}
Consider a population inhabiting two stations; hence, the population consists of two
subpopulations. Any movements of individuals within a station must be neglected; thus,
only the transfer from station to station is considered as a migration act. No spatial
effects in the population dynamics are presumed, for each subpopulation, as soon as no
migration occurs. We shall consider a discrete time model; continuous modelling is
possible, as well, but it brings no new issues but the serious technical difficulties.

Further, the dynamics of each subpopulation is supposed to follow the Verchulst's
equation \cite{verh,sharkov1,sharkov,cyc,progress}. Namely, let $N_t$ and $M_t$ denote
the abundance of the first subpopulation (of the second one, respectively), so that
\begin{subequations}\label{eq:1}
\begin{equation}\label{eq:1-1}
N_{t+1} = a \cdot N_t - b \cdot N_t^2
\end{equation}
\textrm{and}
\begin{equation}\label{eq:1-2}
M_{t+1} = c \cdot M_t - d \cdot M_t^2\;,
\end{equation}
\end{subequations}
respectively, for the case of the absence of migration. We shall implement the migration
into consideration later. Here $a$ and $c$ represent fertility of the relevant
subpopulation, while $b$ and $d$ show the effect of density dependent competition within
a subpopulation, each. It must be said, that unlike for the Verchulst's equation of a
single population, a student may not change the equations (\ref{eq:1-1}, \ref{eq:1-2})
for the dimensionless form, eliminating the coefficients $b$ and $d$. The point is that
the migration effects would break down such transformation. The functions
\begin{subequations}\label{eq:2}
\begin{equation}\label{eq:2-1}
k_r(N_t) = a - b \cdot N_t
\end{equation}
\textrm{and}
\begin{equation}\label{eq:2-2}
k_l(M_t) = c - d \cdot M_t
\end{equation}
\end{subequations}
are the reproduction coefficients, in relevant stations, respectively.

Migration is a transfer of a part of subpopulation from one station into the other.
Migration itself affects the reproduction, survival and other vital functions of an
organism. All these issues will be integrated into the parameter called the {\sl cost of
migration} $\mu$: no negative impact on the reproduction, survival and other vital
functions of an organism is observed, as $\mu = 0$. Otherwise, the growth of $\mu$ yields
a decay in survival of organisms. Rather simple way to introduce the migration cost is
proposed in \cite{zob89,monit89,moya}: $$p = \exp\{- \,\mu\}\;,$$ so that $p$, $0 < p
\leq 1$ becomes a probability of the transfer from station to station with no damages for
further reproduction.

Migration starts, as soon as the conditions of life (measured in the units of
reproduction coefficient (\ref{eq:2})) ``there'' becomes better, than ``here''. This
narrative issue could be easily transformed into the formal way. Migration from
$N$-station (from $M$-station, respectively) starts, if $p \cdot k_l(M) > k_r(N)$ (if
$k_l(M) < p \cdot k_r(N)$, respectively) for the given time moment $t$, $t = 1, 2, 3,
\ldots\,$. We shall suppose, the migration runs as a single act, entirely, and the
reproduction always follows the migration. Thus, a life cycle of a population consists of
two stages: the former is a migration (if it takes place at the given time moment $t$),
and the latter is the reproduction ran according to (\ref{eq:1}), with the abundances
appeared due to migration, at each station, independently.

Schematically, the model works in three steps. \hfil \hfil  \hfil \linebreak \textbf{The
first step} consists in the determination of the fact of migration, i.e. in the
comparison of reproduction coefficients ``there'' and ``here'':
\begin{itemize}
\item if $p \cdot k_l(M) > k_r(N)$, then the migration runs from $N$-station to
$M$-station;
\item if $k_l(M) < p \cdot k_r(N)$, then the migration runs from $M$-station
to $N$-station;
\item otherwise no migration takes place.
\end{itemize}
\textbf{The second step} consists in the determination of the migration flux. Migration
yields the change of life conditions measured in the units of reproduction coefficient
(\ref{eq:2}). An emigration results in the growth of the coefficient, reciprocally,
immigration results in the decrease of that latter, since the coefficients (\ref{eq:2})
are supposed to be a linear descending function of the abundance $N$ (or $M$). The number
of migrating individuals tends to equalize the reproductions coefficients, due to an
abundance redistribution: $$k_l(M - \Delta) = p \cdot k_r(N+ p \cdot \Delta)$$ for the
case of migration from $M$-station to $N$-station, and vice versa: $$p \cdot k_l(M+
p\cdot \Delta) = k_r(N- \Delta)\;.$$ Here the term $p\cdot \Delta$ represents the fact of
mortality of individuals at the course of migration; this is the way to account the
transfer cost, for this model. The migration flux, then, is determined according to
\begin{subequations}\label{eq:3}
\begin{equation}\label{eq:3-1}
\Delta_{MN} = \frac{pc - a +bN - pdM}{b+p^2d}
\end{equation}
\textrm{for the migration from $N$-station to $M$-station, and}
\begin{equation}\label{eq:3-2}
\Delta_{NM} = \frac{pa - c + dM - pbN}{d + p^2b}
\end{equation}
\end{subequations}
for the inverse migration.

\noindent Finally, \textbf{the third step} consists in the reproduction of organisms in
both subpopulations independently, with respect to the abundance ($\widetilde{N}_t$, or
$\widetilde{M}_t$, respectively) resulted from the migration:
\begin{subequations}\label{eq:4}
\begin{equation}\label{eq:4-1}
N_{t+1} = a \cdot \widetilde{N}_t - c \cdot \widetilde{N}_t^2
\end{equation}
\textrm{and}
\begin{equation}\label{eq:4-2}
M_{t+1} = c \cdot \widetilde{M}_t - d \cdot \widetilde{M}_t^2\;.
\end{equation}
\end{subequations}

It should be stressed, that a reproduction in a station (i.e., at the subpopulations)
runs independently, with respect to the abundances occurred due to the migration. If no
migration takes place for some time moment $t$, then $\widetilde{N}_t = N_t$
($\widetilde{M}_t = M_t$, respectively).

\section{Some properties of the model}
\begin{figure*}
\includegraphics{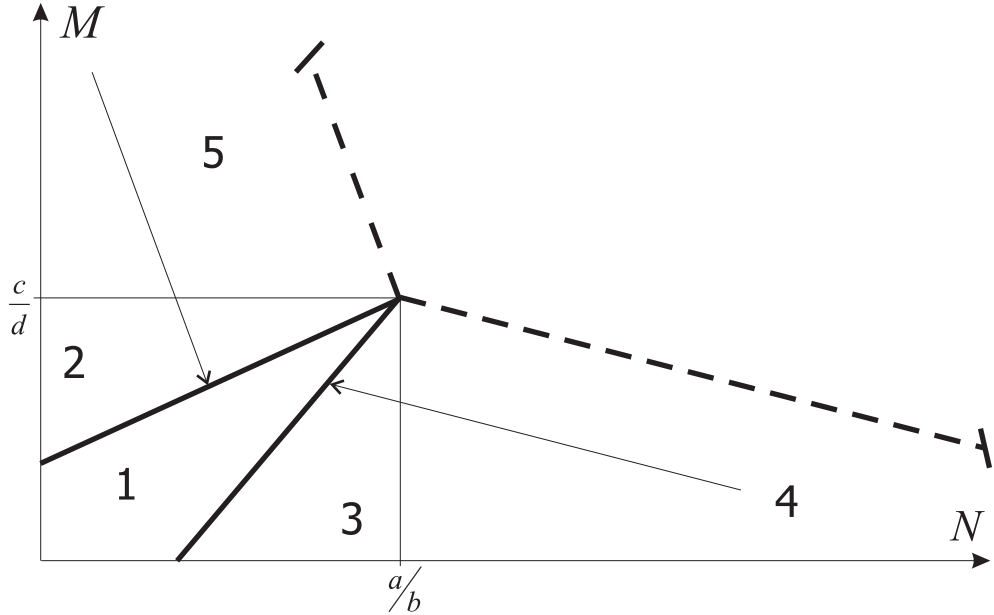}
\caption{\label{f1}Space available for the dynamic trajectories of the model, in case of
non-zero migration; $\mathsf{1}$ -- no migration area, $\mathsf{2, 5}$ -- areas of
migration from $N$-station to $M$-station, $\mathsf{3, 4}$ -- areas of migration from
$M$-station to $N$-station. Horizontal axis shows the abundance at $N$-station, vertical
one shows similar abundance in $M$-station.}
\end{figure*}

The model (\ref{eq:1}~--~\ref{eq:4}) of the smart migration exhibits various dynamic
properties, not observed at the similar population dynamics models with no migration. The
dynamics of two (independent) subpopulations runs inside the rectangular $[0, a/b] \times
[0, c/d]$, if no migration occurs. As soon as $p
> 0$, the dynamics runs at the dovetail shown in Fig.\ref{f1}.
If a trajectory (i.e. a point representing a couple of abundances $(N_t, M_t)$) reaches
the area $\mathsf{1}$, then no migration occurs at that time moment $t$, and no migration
will take place, while the trajectory remains at this area. The area is cut off with two
solid bold lines. These are determined by the equations $$\Delta_{MN} = 0 \qquad
\textrm{and} \qquad \Delta_{NM} = 0\;.$$ Areas $\mathsf{2}$ and $\mathsf{3}$ are the
space dynamically available by each subpopulation, independently, when no migration takes
place.

On the contrary, the migration expands the dynamically reachable area. Such expansion
results from the smart migration: individuals emigrate from the overpopulated station
improving the survival of the entire population. Areas $\mathsf{2}$ and $\mathsf{5}$ in
Fig.\ref{f1} show this expansion. For any point $(N_t, M_t) \in \Omega$, where $\Omega$
is a union of these two areas, the migration gives the projection of this point on the
bold solid line bordering the area $\mathsf{1}$, parallel to the solid dashed line shown
in this Figure. Thin arrow in this Figure illustrates such projection. Similarly, if a
point $(N_t, M_t) \in \Phi$ belongs to the union $\Phi$ of areas $\mathsf{3}$ and
$\mathsf{4}$, then the migration maps it on the (lower) solid bold line bordering area
$\mathsf{1}$. Similarly, it is executed in parallel to the thin arrow, shown in the
figure.

The tangent of the dashed line bordering the area $\mathsf{5}$ is equal $-p^{-1}$;
similarly, for the area $\mathsf{4}$ the tangent is equal to $-p$. The area $\mathsf{1}$
expands, as $p \rightarrow 0$ occupying the entire rectangular $[0, a/b] \times [0,
c/d]$, for $p = 0$. The areas $\mathsf{4}$ and $\mathsf{5}$ become a (semi-infinite)
strip each, of the permanent width. This fact differs the situation of the complete
absence of a migration from the infinitely big migration cost $\mu$. This area collapses
into the line (into the intercept, to be exact) defined the the equation $$bN - dM + a -
c =0\;,$$ as $p \rightarrow 1$.

Next, the migration expands the allowed parameter values. The parameters $a$ and $c$ must
meet the constraint $$1 \leq a \leq 4 \qquad \textrm{and} \qquad 1 \leq c \leq 4\;,$$ in
migration free models \cite{lande,sharkov,sharkov1,ecol2003,2,3,progress}. The equation
(\ref{eq:1-1}) (or (\ref{eq:1-2}), respectively) exhibits elimination of a population, as
$a$ ($c$, respectively) exceeds $4$: maximum of $aN_t - bN_t^2$ may be greater than
$a/b$. The non-zero migration expands the range of the parameter, provided that the
overpopulation in one station will be compensated by the emigration into another.

Equation (\ref{eq:1-1}) (or (\ref{eq:1-2}), respectively) shows various dynamic patterns,
for various parameter $a$ (or parameter $c$, respectively) value. The diversity of limit
regimes of (\ref{eq:1}) varies from a stable fixed point to a strange attractor as a
limit manifold. Similar diversity of regimes could be found at the equation system
(\ref{eq:1-1}, \ref{eq:1-2}), when the migration occurs. In addition, the system
(\ref{eq:1}~--~\ref{eq:4}) exhibits some other regimes, that are not possible with no
migration.

Migration provides a redistribution of individuals over a space. The redistribution is
not random; it results in maximization of the average (over two stations) reproduction
coefficient $k_r(\widetilde{N}) + k_l(\widetilde{M})$. The reproduction coefficient
reaches the maximum in one step; this point results both from the global information
accessibility of the data concerning the environmental conditions (coefficients $a$, $b$,
$c$ and $d$), population density (these are $N_t$ and $M_t$ in both stations, at every
time moment $t$), and the transfer cost $\mu$ (or the probability of the successful
transfer $p$). If $0 < p < 1$, the migration results a decrease of a general abundance,
since a part of individuals (namely, $p\cdot \Delta$) is eliminated, at each time step
$t$. Migration yields no elimination of individuals, as $p=1$.

Various limit regimes occurred due to smart migration effect may be observed. There
exists the stable permanent one-side directed flux of individuals. For example, if $c$ is
big enough, and simultaneously $b$ is sufficiently small (thus increasing the
environmental capacity of $N$-station), one may observe the permanent one-way migration
flux limit regime. Indeed, such regime is determined by the equation
\begin{equation}\label{onemig}
\left\{
\begin{array}{c}
N^{\ast} = a\cdot \left(N^{\ast} + p\Delta^{\ast}\right) - b\cdot \left(N^{\ast} +
p\Delta^{\ast}\right)^2\\
M^{\ast} = c\cdot \left(M^{\ast} - \Delta^{\ast}\right) - d\cdot \left(M^{\ast} -
\Delta^{\ast}\right)^2\;,
\end{array}
\right.
\end{equation}
with $\Delta^{\ast}$ determined by (\ref{eq:3-2}): $$\Delta^{\ast} = \frac{pa - c +
dM^{\ast} - pbN^{\ast}}{d + p^2b}\,.$$ To figure out the impact of the smart migration on
the population dynamics, let's consider a particular case of $b = d = 1$, and $p =1$.
Such choice of the parameters means that the subpopulations differ in their growth rate,
only, and no losses of an abundance are resulted from a migration. The equality $p = 1$
also means a supreme mobility of an individual. Here $\Delta$ becomes equal to $$\Delta =
a-c+M-N\;,$$ and the system (\ref{onemig}) changes for
\begin{equation}\label{onemig1}
\left\{
\begin{array}{rcl}
N & = & a\cdot \left(M + \lambda\right) - \left(M + \lambda\right)^2\\
M & = & c\cdot \left(N - \lambda\right) - \left(N - \lambda\right)^2\;,
\end{array}
\right.
\end{equation}
with $\lambda = a - c$. Here the asterisks are omitted, since it makes no confusion. This
is the system of two polynomials of two variables $\left(M, N\right)$ of (formally) power
$2$ each. To solve a system of two polynomials of two variables, one must develop the
resultant of the system. This former is the determinant of the following matrix:
\begin{widetext}
\begin{equation}\label{detr1}
\left(
\begin{array}{cccc}
0 & 1 & - \left[a(M+\lambda) - (M+\lambda)^2 \right] & 0\\
0 & 0 & 1 & - \left[a(M+\lambda) - (M+\lambda)^2 \right]\\
1 & -(2\lambda + c) & \lambda^2 + c\lambda + M & 0\\
0 & 1 & -(2\lambda + c) & \lambda^2 + c\lambda + M\\
\end{array}
\right)
\end{equation}
\end{widetext}
for the case (\ref{onemig1}). Some roots of (\ref{detr1}) are the solution
of (\ref{onemig}). It should be stressed, that the first term at the first line of the
matrix (\ref{detr1}) becomes non-zero, for the case $b \neq d$.

Suppose, $(N^{\ast},\ M^{\ast})$ is the solution of (\ref{onemig1}), and $N^{\ast} > 0$,
$M^{\ast}> 0$. Here the question arises, whether this solution is stable. The answer on
this question could be obtained due to linear approximation analysis. Let $N_t = N +
\nu_t$ and $M_t = M + \mu_t$, where $\nu_t, \mu_t$ are the small corrections;
substituting such $(N_t,\ M_t)$ into (\ref{onemig}) and omitting the terms of the second
and higher orders, one gets the following matrix for linear approximation:
\begin{equation}\label{ust1}
\left(
\begin{array}{cc}
0 & -2\cdot \left(M+\lambda\right)\\
-2\cdot \left(N-\lambda\right) & 0\\
\end{array}
\right)\;.
\end{equation}
The eigenvalues of matrix (\ref{ust1}) are less, than $1$, when $\left|
\left(M+\lambda\right) \cdot \left(N-\lambda\right) \right| < 1/4$.

Similarly, numerous other limit regimes of different complexity and structure could be
found, as well. One hardly can figure them out explicitly; moreover, there is no much
sense in detailed determination of these complex regimes. Computer simulation makes them
rather obvious.

Here we present the simplest model of a smart migration, strongly opposing the ``reaction
-- diffusion'' methodology. The model is rather clear and apparent. The simplicity
results from the peculiar feature of the model; this is the case of globally informed
individuals. Indeed, an individual makes a decision whether it would migrate, or would
stay in the station, referring to the available information concerning the life
conditions. So, the key issue is what information towards that matter is available.

Let's concentrate on some mathematical issues followed from the smart migration. The
first one is that the dynamics of a population with smart migration is irreversible in
time. Indeed, the migration at the model (\ref{eq:1}~--~\ref{eq:4}) is a projection, from
mathematical point of view; thus, a set of different states are transformed into unique
one, and there is no regular way to figure out which one was preceding the observed
abundance resulted from the migration act.

Another important issue is that it expands both permissible phase space, and the
parameters values, in comparison to similar models with no migration. It should be said,
that such effects, probably, also could be observed (at least, for some peculiar
combinations of the parameters) for the systems with random, or aimless migration. We
doubt that the direct comparison of the areas of permissible phase variables, or the
parameters, for the case of smart vs. the aimless migration makes much sense. Obviously,
the model is rather simple and specific to pretend to describe properly any real
biological system. The specificity here manifests in the global information
accessibility; in occupation not more than two stations; in the absence of
``microscopic'' consideration of dynamics within a station.

The main purpose of this paper is to show the methodology of the modelling of spatially
distributed populations (and other biological communities) with no chemical analogies,
rather than to find out some peculiar dynamic regime pretending to match exactly a
dynamics of a real biological community.

The model provides that the life conditions at the residence station are converted into a
couple of parameters $a$ and $b$ ($c$ and $d$, respectively). Besides, it is supposed
that the density of the subpopulation (or its total abundance $N$, or $M$, respectively)
is known, as well. Not discussing at the moment the details of the detection of
population density, or other conditional parameters {\sl per se}, suppose that the
environmental abundance (density, indeed) and other parameters are detectable for an
individual. Besides, the model (\ref{eq:1}~--~\ref{eq:4}) suggests that similar
parameters, and the abundance are known, at the distant station. Such suggestion makes
the individuals to be globally informed. A feasibility of such presumption is doubtful,
nevertheless, this assumption is a common place for mathematical population biology
\cite{levin97,ecol2003,lande,ber3,progress,edel2,alstad}.

The methodology presented above is rather powerful, and provides a researcher with the
tool for studying spatially distributed populations with no artificial and absolutely
unrealistic hypotheses towards the microscopic behavior of individuals, i.e. the
randomness and aimlessness of their transfers over a space. One sees the following
furthering of the approach described here. First, a two-species (or several species)
communities could be described within the framework of the methodology. Again, one should
consider a two-station model, where each species (say, predators and preys) migrate from
station to station and back. A dynamics within a station might be modelled with the the
most common equation (say, with Lotka-Volterra equation), thus explicating the effect of
smart migration in the dynamics of a multi-species community.

Both single-species, and multi-species models with smart migration may incorporate
various patterns of information accessibility, for individuals. The model presented above
is based on the hypothesis of the global accessibility of information to an individual.
It means, that an individual knows the conditions of life (expressed in the coefficients,
at the case of the model (\ref{eq:1} -- \ref{eq:4}), both at the station of residence,
and the station of immigration; it knows the density of each subpopulation. Finally, the
an individual knows the transfer cost, in this case.

The hypothesis of total lack of information available to individuals opposes the idea
mentioned above. Here an individual operates with the inner, extremely local information,
when makes a decision on the change of the location for another one. In brief, such
situation could be described like a threshold migration, where the transfer act takes
place only when the local conditions become worse than some individually defined level.
Still, the situation of the total lack of information is not equivalent to random and
aimless migration. The difference becomes clear, if one considers the situation of the
transfer act occurrence, while the local density at the occupancy place is still very
low. The smart migration under the total lock of information would start up, while
chemically-like diffusive migration will not take place, in such case.

All these assumptions seem to be too strong and specific. In general, the individuals
operate with a part of information. There are several problems here, both of mathematical
origin, and of biological essence. The first one consists in exact and comprehensive
definition of what exactly is known to individuals. Next one is the discretion between
the behavioral patterns supported by the reasonable choice of the way to behave, and
those determined instinctively. Consider the seasonal bird migration. Surely, the fact
that some species change a reproduction site for a winter spending site, falls beyond the
will of a bird, it does not make a matter of reasonable of self-made choice: that is the
instinct forcing them to fly away. On the other hand, the choice of a peculiar site to
spend a night (if any) is made by the birds in a flock reasonably, with respect to the
detail features and circumstances of the current situation. Finally, the problem arises
when one tends to determine how far (in space) the individuals are able to collect and
process the information (concerning the living conditions ``there'').

The model (\ref{eq:1}~--~\ref{eq:4}) implies that the transfer cost $\mu$ is symmetrical,
and does not depend on the direction of migration. It might be so, while more realistic
idea is that the transfer cost should be unsymmetrical. Evidently, the simplest way to
figure out the transfer cost is to split it into three parts:
\begin{equation}\label{eq:9}
\mu = \mu_{\textrm{out}}(\mathsf{A}) + \mu_{\textrm{in}}(\mathsf{B}) +
\delta(\mathsf{A},\mathsf{B})\;.
\end{equation}
Here $\mu_{\textrm{out}}(\mathsf{A})$ is the transfer cost of successful emigration from
the station~$\mathsf{A}$; $\mu_{\textrm{in}}(\mathsf{B})$ is the transfer cost of
successful introduction into the station~$\mathsf{B}$, respectively; finally,
$\delta(\mathsf{A},\mathsf{B})$ is the pure transfer cost from station~$\mathsf{A}$ to
station~$\mathsf{B}$. Obviously, one should expect to face the asymmetry
$$\mu_{\textrm{out}}(\mathsf{A}) \neq \mu_{\textrm{out}}(\mathsf{B}) \qquad \textrm{and}
\qquad \mu_{\textrm{in}}(\mathsf{A}) \neq \mu_{\textrm{in}}(\mathsf{B})\;,$$ in general.
Also, a symmetry of pure transfer cost $\delta(\mathsf{A},\mathsf{B}) =
\delta(\mathsf{B},\mathsf{A})$ is doubtful. It should be said, that this point makes no
problem in its implementation at the model (\ref{eq:1}~--~\ref{eq:4}).

Another significant constraint of the model (\ref{eq:1}~--~\ref{eq:4}) is the spatial
structure limited with two stations. Indeed, an expansion of the approach presented above
for the case of several stations, and, ultimately, for a continuous, or quasi-continuous
case of a habitat is strongly desirable. Suppose, a population inhabits three stations;
here we presume the global information accessibility, as well. Suppose, further, the
conditions (i.e., the abundances and the parameters) make the situation when the
individuals from the station $\mathsf{A}$ must migrate either to station $\mathsf{B}$, or
to station $\mathsf{C}$. No one knows exactly, in advance, what is a proportion of
individuals immigrating into the station $\mathsf{B}$ vs. those immigrating the station
$\mathsf{C}$. This is the main obstacle here. There exists the approach withdrawing this
discrepancy; it is based on the interval mathematics \cite{int,int1,int2,int3,int4}. The
detailed discussion of that issue falls beyond the scope of this paper.

\section{Conclusion}
The model described above implements the methodology of evolution optimality into the
problem of the modelling of spatially distributed populations. Migration causes the
growth on net reproduction (which is a reproduction rate, in our case), in average, over
the space. The model comprises the simplest case of two stations, where the spatial
distribution is restricted to a transfer of individuals from station to station and back;
the transfer cost is supposed to be symmetrical one. The model shows the expansion of the
environment capacity, in comparison to the case of the migration absence.

\begin{acknowledgments}
I am thankful to Prof. Alexander Gorban from Liechester University for the general
presentation of the problem, and for permanent cooperation. The work was partially
supported be Krasnoyarsk science foundation, grant 6F050C.
\end{acknowledgments}


\begin{thebibliography}{99}

\bibitem{alstad} Alstad~D. (2001) Basic Populus Models of Ecology, Prentice
Hall.

\bibitem{bern} Bernstein~R. Population Ecology: An Introduction to
Computer Simulations. Wiley \& Sons.

\bibitem{ber} Berryman~A.A. (2002) Population cycles: the case for
trophic interactions. New York: Oxford University Press, 321 p.

\bibitem{ber3} Berryman~A.A. (2003) On principles, laws and theory in
population ecology // \textit{Oikos}, \textbf{103}, pp.695 -- 701.

\bibitem{cyc} Castro-e-Silva~A., Bernardes~A.T. (2001) Analysis of chaotic
behaviour in the population dynamics // \textit{Physica}~\textbf{A 301}, pp.63 -- 70.

\bibitem{sci2000} Condit~R., Ashton~P.S., Baker~P.,
Bunyavejchewin~S., Gunatilleke~S., Gunatilleke~N., Hubbell~S.P., Foster~R.B., Itoh~A.,
LaFrankie~J.V., Hua Seng Lee, Losos~E., Manokaran~N., Sukumar~R., Yamakura~T. (2000)
Spatial patterns in the distribution of tropical tree species // \textit{Science}
\textbf{288}, pp.1414 -- 1418.

\bibitem{edel2} Edelstein~L., Edelstein~K. (1988)
Mathematical Models in Biology. Birkheauser Mathematics Series.

\bibitem{ezer} Ezersky~A.B., Rabinovich~M.I. (1990) Nonlinear-wave
competition and anisotropic spectra of spatiotemporal chaos of Faraday ripples //
\textit{Europhysics Letters}, \textbf{13}(3), pp.243 -- 249.

\bibitem{2} Gilpin~M.E., Ayala~F.J. (1973) Global models of growth
and competition // \textit{Proc.Natl. Acad.Sci.}, \textbf{70}, pp.3590 -- 3593.

\bibitem{gi-co04} Ginzburg~L.R., Colyvan~M. (2004) Ecological
Orbits: how planets move and populations grow. New York: Oxford University Press, 354 p.

\bibitem{3} Gromov~M. (2000) A dynamical model for synchronisation and for
inheritance in microevolution: a survey of papers of A.Gorban, The talk given in the IHES
seminar, ``Initiation to functional genomics: biological, mathematical and algorithmical
aspects'', Institut Henri Poincar\'{e}, November 16, 2000.

\bibitem{otbor3} Gorban~A.N. (1984) Equilibrium encircling. Equations of
chemical kinetics and their thermodynamic analysis, Novosibirsk: Nauka plc., 1984.

\bibitem{otbor1} Gorban~A.N. (1992) Dynamical systems with inheritance, In:
Some problems of community dynamics, R.G. Khlebopros (ed.); Novosibirsk: Nauka plc.,
pp.40 -- 72.

\bibitem{otbor4} Gorban~A.N. (2005) Systems with inheritance: dynamics of
distributions with conservation of support, natural selection and finite-dimensional
asymptotics // arXiv:cond-mat/0405451

\bibitem{na2} Gorban~A.N., Karlin~I.V. (2003) Family of additive
entropy functions out of thermodynamic limit // Physical Review~\textbf{E 67}, 016104.

\bibitem{na4} Gorban~A.N., Khlebopros~R.G. (1988) Demon of
Darwin: Idea of optimality and natural selection, Moscow: Nauka (FizMatGiz).

\bibitem{bolgar} Gorban~A.N., Sadovsky~M.G. (1987) Population
mechanisms of cell aggregation in continuous cultivation systems // \textit{Biotechnology
and Biotechnique}, \textbf{2}(5), pp.34 -- 36.

\bibitem{zob89} Gorban~A.N., Sadovsky~M.G. (1989) Optimization
models of spatially distributed populations: Alle's effect // \textit{Rus.J.General
Biol.}, \textbf{50}(1), pp.66 -- 72.

\bibitem{monit89} Gorban~A.N., Sadovsky~M.G. (1989)
Optimization models: the case of globally informed individuals. In: Problems of
environmetnal monitoring and modelling of ecosystems, vol.\textbf{11}, Leningrad:
Gidrometeoizdat, pp.198 -- 203.

\bibitem{haldane} Haldane~J.B.S. (1990) The Causes of Evolution, Princeton
Science Library, Princeton University Press.

\bibitem{int} Jaulin~L., Kieffer~M., Didrit~O., Walter~E. (2001)
Applied Interval Analysis, with Examples in Parameter and State Estimation, Robust
Control and Robotics. Heidelberg, Hamburg: Springer-Verlag.

\bibitem{kolmog} Kolmogorov~A.N., Petrowsky~I.G., Piscounov~N.A.
(1937) \'{E}tude de l'\'{e}quation de la diffusion avec croissance de la quantit\'{e} de
mati\`{e}re et son application \`{a} un probl\`{e}me biologique //
\textit{Mosc.Univ.Bull.Math.} \textbf{1}, pp.1 -- 25.

\bibitem{int2} Kr\"{a}mer~W., von~Gudenberg~J.W. (2001)
Scientific Computing, Validated Numerics, Interval Methods, Kluwer,
Boston/Dordrecht/London.

\bibitem{int3} Kulisch~U.W. (2002) Advanced Arithmetic for the Digital
Computer, Springer-Verlag, Wien.

\bibitem{lande} Lande~R., Engen~S., Saether~B.-E., Sther~B.-E.
(2003) Stochastic Population Dynamics in Ecology and Conservation (Oxford Series in
Ecology and Evolution), Oxford Univ.Press.

\bibitem{ecol2003} Law~R., Murrell~D.J., Dieckmann~U. (2003) Population
growth in space and time: spatial logistic equations // \textit{Ecology}, \textbf{84}(1),
pp.252 -- 262.

\bibitem{levin97} Levin~S.A., Grenfell~B., Hastings~A.,
Perelson~A.S. (1997) Mathematical and Computational Challenges in Population Biology and
Ecosystems Science // \textit{Science}, \textbf{275}, pp.334 -- 43.

\bibitem{levitt} Levitt~P.R. (1975) General kin
selection models for genetic evolution of sib altruism in diploid and haplodiploid
species // \textit{Proc.Nat.Acad.Sci. USA}, \textbf{72}(11), pp.4531 -- 4535.

\bibitem{lotka} Lotka~A.J. (1925) Elements of Physical Biology.
Baltimore: Williams \& Wilkens.

\bibitem{lvov} L'vov~V.S. (1994) Wave turbulence under parametric excitation
applications to magnets, Springer, Berlin, Heidelberg.

\bibitem{insect} Maron~J.L., Harrison~S. (1997) Spatial Pattern
Formation in an Insect Host-Parasitoid System // \textit{Science}, \textbf{278}, pp.1619
 -- 1621.

\bibitem{progress} Matsuda~H.N., Ogita~A., Sasaki~A., Sat\~{o}~K. 1992.
Statistical mechanics of population: the lattice Lotka-Volterra model // \textit{Progress
in Theoretical Physics}, \textbf{88}, pp.1035 -- 1049.

\bibitem{murray} Murray~J.D. (2002) Mathematical Biology, vols.
\textbf{I--II}, Third edition. Berlin: Springer-Verlag.

\bibitem{int4} Neumaier~A. (2001) Introduction to Numerical Analysis.
Cambridge: Cambridge Univ. Press.

\bibitem{int1} Petkovic~M.S., Petkovic~L.D. (1998) Complex
Interval Arithmetic and Its Applications, John Wiley.

\bibitem{rozonoer} Rozonoer~L.I., Sedyh~E.I. On the mechanisms of
evolution of self-reproduction systems, $\mathsf{1}$, \textit{Automation and Remote
Control}, \textbf{40}(2), pp.243 -– 251; $\mathsf{2}$, ibid., \textbf{40}(3), pp.419 --
429; $\mathsf{3}$, ibid, \textbf{40}(5), pp.741 -- 749.

\bibitem{zelkniga} Sadovsky~M.G., Gurevich~Yu.L., Manukovsky~N.S.
(1989) Kinetics of cell aggregation in continuous cultivation. In: Dynamics of chemical
and biological systems, Novosibirsk: Nauka plc, pp.134 -- 158.

\bibitem{moya} Sadovsky~M.G. (1992) Optimization modelling
of globally informed individuals. In: Mathematical modelling in biology and chemistry.
Evolution approach, Novosibirsk: Nauka plc, pp.36 -- 67.

\bibitem{otbor2} Semevsky~F.N., Semenov~S.M. (1984.)
Mathematical modelling of ecological processes. Leningrad: Gidrometeoizdat, 426 p.

\bibitem{sharkov1} Sharkovsky~A.N. (1965) On cycles and the structure of
continuous mapping // \textit{Ukranian mathematical journal}, \textbf{17}(3), pp.104 --
111.

\bibitem{sharkov} Sharkovsky~A.N. (1983) Difference equations and
population dynamics / In: Mathematical methds in biology. Proc.$2^{\textrm{nd}}$ Ukranian
Conf., Kiev: Naukova Dumka plc., pp.143 -- 156.

\bibitem{xehi} Smitalova~K., Sujan~S. (1991) A mathematical
treatment of dynamical models in biological science. Ellis Horwood.

\bibitem{annals} Soares~P., Tom\'{e}~M. (1999) Distance-dependent
competition measures for eucalyptus plantations in Portugal // \textit{Annals of Forestry
Science}, \textbf{56}, pp.307 -- 319.

\bibitem{ameba} Strassmann~J.E., Yong~Zhu, Queller~D.C. (2000)
Altruism and social cheating in the social amoeba {\sl Dictyostelium discoideum} //
\textit{Nature}, \textbf{408}, pp.965 -- 967.

\bibitem{1} The geometry of ecological interactions:
simplifying spatial complexity. Dieckmann~U., Law~R., Metz~J.A.J., editors. Cambridge
University Press, Cambridge, UK.

\bibitem{thieme} Thieme~H.R. (2003) Mathematics in Population Biology
(Princeton Series in Theoretical and Computational Biology). Princeton Univ.Press.

\bibitem{turch} Turchin~P. (2003) Complex Population Dynamics: A
Theoretical/Empirical Synthesis. Princeton University Press, 357 p.

\bibitem{verh} Verhulst~P.(1845) Recherches math\'{e}matiques sur la loi
d'accroissement de la population // \textit{Nouv.M\'{e}m.de l'Academie Royale des Sci.et
Belles-Lettres de Bruxelles} \textbf{18}, pp.1 -- 41.

\bibitem{volt} Volterra~V. (1926) Variazioni e fluttuazioni del numero
d'individui in specie animali conviventi. \textit{Mem.R.Accad.Naz.dei Lincei}, Ser.VI, 2.

\bibitem{volt1} Volterra~V. (1931) Le\c{c}ons sur la th\'{e}orie
math\`{e}matique de la lutte pour la vie, Gauthier-Villars, Paris.

\bibitem{zah-usp} Zakharov~V.E., L'vov~V.S., Starobinets~S.S. (1974)
Turbulence of spin-waves beyond threshold of their parametric-excitation //
\textit{Uspekhi Fizicheskikh Nauk} \textbf{114}(4), pp.609 -- 654; English translation
\textit{Sov.Phys.-Usp.} \textbf{17}(6) (1975), pp.896 -- 919.

\bibitem{zah-spek} Zakharov~V.E., L'vov~V.S., Falkovich~G.E.
(1992) Kolmogorov spectra of turbulence, vol.\textbf{1} Wave Turbulence. Springer,
Berlin.

\end{thebibliography}
\end{document}